\documentclass[preprint,authoryear,12pt]{elsarticle}

\usepackage{graphics}
\usepackage{graphicx}
\usepackage{epsfig}

\usepackage{amssymb}
\usepackage{amsthm}

\usepackage{lineno}

\begin{document}

\begin{frontmatter}



\title{
Analysis of the turbulent law of the wall through the finite scale Lyapunov theory}


\author{Nicola de Divitiis}

\address{"La Sapienza" University, Dipartimento di Ingegneria Meccanica e Aerospaziale, Via Eudossiana, 18, 00184 Rome, Italy}

\begin{abstract}
This work analyzes the turbulent velocity distribution in proximity of the wall 
using the finite-scale Lyapunov theory just presented in previous works. 
This theory is here applied to the steady boundary layer under the hypothesis 
of moderate pressure gradient and fully developed flow along the streamwise direction.
The analysis gives an equation for the velocities correlation and 
identifies the parameters of the expression of the average velocity 
through the statistical properties of the velocity correlation functions.
In particular, the von K\'arm\'an constant, theoretically calculated, is about 0.4, and the dimensionless Prandtl's length is in function of the Taylor-scale Reynolds number.
The study provides the average velocity distribution and gives also the variation laws
of the other variables, such as Taylor scale and Reynolds stress. 
The obtained results show that the finite-scale Lyapunov theory is adequate
for studying the turbulence in the proximity of the wall.
\end{abstract}


\end{frontmatter}

\newcommand{\no}{\noindent}
\newcommand{\be}{\begin{equation}}
\newcommand{\ee}{\end{equation}}
\newcommand{\bea}{\begin{eqnarray}}
\newcommand{\eea}{\end{eqnarray}}
\newcommand{\bc}{\begin{center}}
\newcommand{\ec}{\end{center}}

\newcommand{\calr}{{\cal R}}
\newcommand{\calv}{{\cal V}}

\newcommand{\bff}{\mbox{\boldmath $f$}}
\newcommand{\bfg}{\mbox{\boldmath $g$}}
\newcommand{\bfh}{\mbox{\boldmath $h$}}
\newcommand{\bfi}{\mbox{\boldmath $i$}}
\newcommand{\bfm}{\mbox{\boldmath $m$}}
\newcommand{\bfp}{\mbox{\boldmath $p$}}
\newcommand{\bfr}{\mbox{\boldmath $r$}}
\newcommand{\bfu}{\mbox{\boldmath $u$}}
\newcommand{\bfv}{\mbox{\boldmath $v$}}
\newcommand{\bfx}{\mbox{\boldmath $x$}}
\newcommand{\bfy}{\mbox{\boldmath $y$}}
\newcommand{\bfw}{\mbox{\boldmath $w$}}
\newcommand{\bfk}{\mbox{\boldmath $\kappa$}}

\newcommand{\bfA}{\mbox{\boldmath $A$}}
\newcommand{\bfD}{\mbox{\boldmath $D$}}
\newcommand{\bfI}{\mbox{\boldmath $I$}}
\newcommand{\bfL}{\mbox{\boldmath $L$}}
\newcommand{\bfM}{\mbox{\boldmath $M$}}
\newcommand{\bfS}{\mbox{\boldmath $S$}}
\newcommand{\bfT}{\mbox{\boldmath $T$}}
\newcommand{\bfU}{\mbox{\boldmath $U$}}
\newcommand{\bfX}{\mbox{\boldmath $X$}}
\newcommand{\bfY}{\mbox{\boldmath $Y$}}
\newcommand{\bfK}{\mbox{\boldmath $K$}}

\newcommand{\bfrho}{\mbox{\boldmath $\rho$}}
\newcommand{\bfchi}{\mbox{\boldmath $\chi$}}
\newcommand{\bfphi}{\mbox{\boldmath $\phi$}}
\newcommand{\bfPhi}{\mbox{\boldmath $\Phi$}}
\newcommand{\bflambda}{\mbox{\boldmath $\lambda$}}
\newcommand{\bfxi}{\mbox{\boldmath $\xi$}}
\newcommand{\bfLambda}{\mbox{\boldmath $\Lambda$}}
\newcommand{\bfPsi}{\mbox{\boldmath $\Psi$}}
\newcommand{\bfomega}{\mbox{\boldmath $\omega$}}
\newcommand{\bfOmega}{\mbox{\boldmath $\Omega$}}
\newcommand{\bfeps}{\mbox{\boldmath $\varepsilon$}}
\newcommand{\bfepsn}{\mbox{\boldmath $\epsilon$}}
\newcommand{\bfzeta}{\mbox{\boldmath $\zeta$}}
\newcommand{\bfkappa}{\mbox{\boldmath $\kappa$}}
\newcommand{\itPsi}{\mbox{\it $\Psi$}}
\newcommand{\itPhi}{\mbox{\it $\Phi$}}
\newcommand{\bint}{\mbox{ \int{a}{b}} }
\newcommand{\ds}{\displaystyle}
\newcommand{\Sum}{\Large \sum}


\section{\bf Introduction}
 \label{intro}

The average velocity law near the wall is well known from many years (\cite{Karman30}) for what concerns the flow in smooth pipes (\cite{Nikuradse}, \cite{Reichardt}) and in the cases of turbulent boundary layers with moderate pressure gradient  (\cite{Klebanoff_1}, \cite{Klebanoff_2}, \cite{Smith}). This law, usually expressed as
\bea
\begin{array}{l@{\hspace{+0.2cm}}l}
\ds U^+ = y^+,  \ \ \     y^+ \le 5, 
\end{array}
\label{wall law 00}
\eea 
\bea
\begin{array}{l@{\hspace{+0.2cm}}l}
\ds U^+ \simeq A \ln y^+ +B, \ \ \ \ \ 5 < y^+ < 30,
\end{array}
\label{wall law 0}
\eea 
\bea
\begin{array}{l@{\hspace{+0.2cm}}l}
\ds U^+ = \frac{1}{k} \ln y^+ + C, \ \ \ \  y^+ > 30,
\end{array}
\label{wall law}
\eea 
gives the average velocity in different regions near the wall,
where  $y^+=y U_T/\nu$ and $U^+ = U/U_T$ are the dimensionless normal coordinate and average velocity, and $U_T = \sqrt{\nu (\partial U/\partial y})_0$ is the friction velocity.

\begin{figure}[t]
	\centering
         \includegraphics[width=0.60\textwidth]{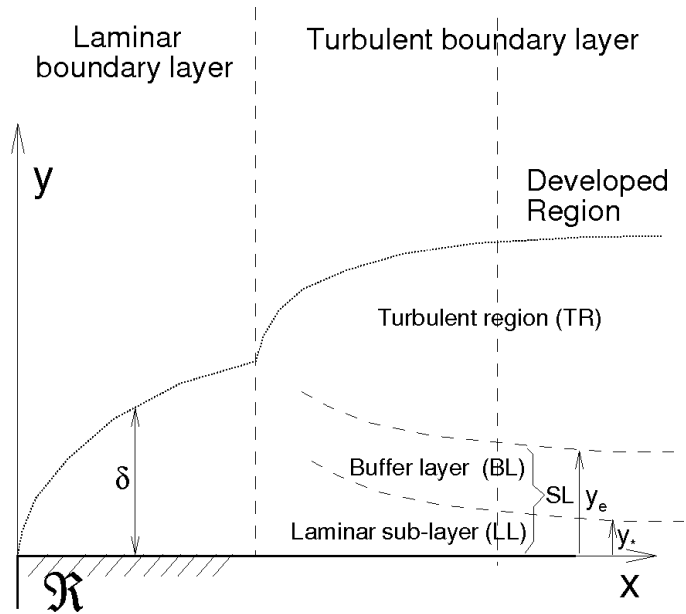}
\caption{Schematic of the boundary layer.}
\label{figura_1}
\end{figure}

These expressions and the values of $A$, $B$, $C$ and $k$, seem to be universal properties of the flow, which do not depend on the Reynolds number. With reference to Fig. \ref{figura_1}, these equations, obtained through considerations of dimensional analysis and of self-similarity, hold under the hypothesis of fully developed parallel flow along the streamwise direction $x$ (\cite{Karman30}).
Specifically, Eq.  (\ref{wall law 00}), being the direct consequence of the wall boundary condition and of the definition of $U_T$, expresses the velocity distribution in the laminar sub-layer LL, a domain adjacent to the wall where the effects of the viscosity are dominant. 
Equation (\ref{wall law}) holds in the turbulent region TR, a zone of non-homogeneous turbulence, where the Taylor scale Reynolds number is high and variable with $y$.
Into Eq. (\ref{wall law}), $k$ is the von K\'arm\'an constant which according to 
\cite{Karman30} and in line with Eq. (\ref{wall law}), can also be expressed as
\bea
\ds  k =  - \lim_{y^+ \rightarrow y^+_e} 
\frac{\ds \frac{d^2 U^+}{d {y^+}^2}}{\ds \left( \frac{ d U^+}{d y^+} \right)^2}
\label{vk alt}
\eea
where $y^+_e$ defines the lower limit of the turbulent region.
As far as Eq. (\ref{wall law 0}) is concerned, this describes the velocity variations in the buffer layer BL, an intermediate zone between TR and LL in which viscous and inertia forces are comparable as order of magnitude.

$k$ and $A$ are directly related to the velocity variations along $y$, whereas $C$ and $B$ express the order of magnitude of $U$ with respect to the friction velocity in case of smooth wall.
These constants are dimensionless free parameters which can not be theoretically calculated (\cite{Landau}), therefore several experiments dealing with their determination were carried out (\cite{Fernholz},  \cite{Zagarola}), and scaling law similarities were proposed to justify Eq. (\ref{wall law}) (\cite{Barenblatt}).
These constants can be also identified through the elaboration of the results
of direct numerical simulations of the Navier-Stokes equations (\cite{Spalart}, 
\cite{Fernholz} and references therein).
From the various sources of the literature, 
$k = 0.36 \div 0.44$, 
 $C = 4.5 \div 7.5$, 
$A \simeq 5 \div 6$ and
$B \simeq -5 \div -3$.

Another law of the wall, also valid under the conditions of parallel flow and self-similarity, and that seems to show properties of universality, pertains the Reynolds stress $\langle u_x u_y \rangle$
\bea
- \langle u v \rangle^+ = g(y^+), \ \ \ 
\langle u v \rangle^+ = \langle u_x u_y \rangle /U_T^2
\label{uv classic}
\eea
This behaves like $g \approx {y^+}^3$ very near the wall, and
at about $y^+ = 5$, exhibits an inflection point whose coordinate provides the
order of magnitude of the laminar sub-layer tickness. 
The value $-\langle u v \rangle^+$ = 0.5, achieved in the buffer layer at about 
$y^+ = 10 \div 13$, corresponds to the maximum turbulent energy production due to the mean flow, whereas far from the wall $g \approx 1$ (\cite{Hinze}, \cite{Tennekes}).

Although all these distributions seem to be universal laws, 
 according to \cite{Landau} and \cite{Fernholz}, there is not an adequate theory based on physical conjectures, other than the direct numerical simulations of the Navier Stokes equations, which leads to the calculation of the free parameters of these laws.
This represents the main motivation of the present work, whose purpose is to apply
the finite scale Lyapunov theory, described in \cite{deDivitiis_3}, to determine the wall velocity distribution and its parameters.

\bigskip

The first part of the work summarizes the main results of \cite{deDivitiis_3}, which deals with the steady homogeneous turbulence with uniform velocity gradient, and then shows that such results can be extended also to the non-homogeneous turbulence with shear rate.
We give a reasonable demonstration that the equation of the correlation function obtained by  \cite{deDivitiis_3} can hold also in the case of non-homogeneous turbulence, in particular in the turbulent zone of the boundary layer.
This result allows to express, in the turbulent region, the several quantities such as 
dimensionless velocity gradient and turbulent kinetic energy
in function of the local Taylor-scale Reynolds number, whereas
in the laminar sub-layer and in the buffer region, adequate variation laws of the 
variables are assumed, which satisfy the wall boundary conditions 
and match the values of the variables in the adjacent domains.

The analysis gives the von K\'arm\'an constant in terms of the variables at
 the edge of the turbulent region, and provides a logarithmic velocity profile in TR, 
different from Eq. (\ref{wall law}), which exhibits free parameters ($k$ included).
The knowledge of the statistical properties of the spanwise correlation function leads to the estimation of the variables at $y^+_e$ and thus to identify $k$ and the other parameters.
Therefore, the von K\'arm\'an is here theoretically obtained, resulting to be about 0.4,
in line with experiments and numerical simulations, and the other results, in agreement with the several data from the literature, show that the finite scale Lyapunov theory can be an adequate tool for studying the wall turbulence.

\bigskip

\section{\bf Resume  }
\label{Resume}

This section summarizes the main results of \cite{deDivitiis_3} which regards the steady homogeneous turbulence in the presence of an average velocity gradient $\nabla_{\bf x} \bf U$.
There, the author, applying the Finite-scale Lyapunov theory (\cite{deDivitiis_1}) and the Liouville theorem, proposes the following evolution equation of the pair distribution function $F^{(2)}$ of fluid velocities in case of arbitrary flow
\bea
\ds \frac{\partial F^{(2)} }{\partial t}  
+ \nabla_{{\bf x}}  F^{(2)} \cdot {\bf v} 
+ \nabla_{{\bf x}'}  F^{(2)} \cdot {\bf v}'  
= \lambda(r) \left(  F^{(2)}_0 -  F^{(2)}  \right) -J_D   
\label{3}
\eea 
where $F^{(2)}_0$ is the pair distribution function of the isotropic turbulence which exhibits the same momentum and kinetic energy of $F^{(2)}$, 
whereas $-J_D$ represents the rate of $F^{(2)}$ caused by the rate of the turbulent kinetic energy.
$\lambda (r)$ is the finite-scale Lyapunov exponent associated to the finite-scale 
$r = \vert {\bf x}'-{\bf x} \vert$, 
and $\bf v$ and $\bf v'$ are the fluid velocities calculated at $\bf x$ and $\bf x'$ respectively.
In case of homogeneous turbulence, \cite{deDivitiis_3} shows that the steady distribution function reasonably tends to a quantity which depends upon $F^{(2)}_0$, $\nabla_{\bf x} {\bf U}$ and $\lambda$
\bea
\ds F^{(2)} = F^{(2)}_0 + \frac{1}{\lambda(r)}
\left(
\frac{\partial F^{(2)}_0}{\partial v_j}  \frac{\partial U_j}{\partial x_p} v_p 
+ \frac{\partial F^{(2)}_0}{\partial v'_j}  \frac{\partial U_j}{\partial x_p} v'_p 
\right) 
\label{F2}
\eea
The velocity correlation tensor $R_{k i} =\langle u_i u_j' \rangle$ is then calculated, by definition 
\bea
\begin{array}{c@{\hspace{+0.2cm}}l}
\ds R_{k i} = \int_v \int_{v'} F^{(2)} u_k u'_i \  d^3u \ d^3u' =  
R_{k i 0} - \frac{1}{\lambda} 
\left( \frac{\partial U_k}{\partial x_p} R_{p i 0} 
+  \frac{\partial U_i}{\partial x_q} R_{k q 0} 
 \right) 
\end{array}
\label{R_0}
\label{R}
\eea
where $R_{k i 0}$ is the second order velocity correlation tensor associated to 
the isotropic turbulence (\cite{Karman38}, \cite{Batchelor53}) 
\bea
R_{k i 0} ({\bf r}) = u^2 \left( (f -g) \frac{r_k r_i}{r^2} + g \delta_{k i} \right) 
\label{R0}
\eea
being $f$ and $g = f + 1/2 \ r \ \partial f/\partial r$ longitudinal and lateral
velocity correlation functions, respectively, and ${\bf u} = {\bf v} -{\bf U}$ is the 
fluctuating velocity.
For $r = 0$, Eq. (\ref{R}) provides the expression of the Reynolds stresses
in function of $\nabla_{\bf x} {\bf U}$
\bea
\left\langle  u_k u_i \right\rangle = u^2 \left( \delta_{k i} 
-\frac{1}{\Lambda} \left( \frac{\partial U_k}{\partial x_i}   
+  \frac{\partial U_i}{\partial x_k}    \right)  \right) 
\label{Boussinesq}
\label{u_x}
\eea 
Equation (\ref{Boussinesq}) is a Boussinesq closure of the Reynolds stress, being 
$u = \langle u_i u_i\rangle /3$, $\left\langle . \right\rangle$ indicates the average calculated on the ensamble of the fluid velocity, and $\Lambda= \lambda(0)$ is the maximal Lyapunov exponent.
The Lyapunov theory presented in \cite{deDivitiis_1} shows that
$\Lambda = u/\lambda_T$, being $\lambda_T$ the Taylor scale. 

Next, the condition of steady flow leads to the following ordinary differential equation
for $f$ \cite{deDivitiis_3}
\bea
\begin{array}{l@{\hspace{-0.cm}}l}
\ds  \sqrt{\frac{1-f}{2}} \ \frac{d f} {d \hat{r}}  +
\ds \frac{2}{R_T}  \left(  \frac{d^2 f} {d \hat{r}^2} +
\ds \frac{4}{\hat{r}} \frac{d f}{d \hat{r}}  \right) + \frac{10}{R_T} 
\frac{f \ \hat{r}}{\sqrt{2(1-f)}} = 0
\end{array}
\label{I2 steady}  
\eea
whose boundary conditions can be reduced to the following conditions of $f$
in the origin $r=0$
\bea 
\ds f(0)=1, \ \ \ \frac{d f (0)} {d \hat{r}} = 0
\label{bc3}
\eea
where $\hat{r} = r/\lambda_T$,  $d^2f/d\hat{r}^2 (0)\equiv -1$, 
and $R_T = u \lambda_T /\nu $ is the Taylor scale Reynolds number. 
Thus, Eqs. (\ref{I2 steady}) and (\ref{bc3}) express an initial condition problem
whose initial condition is given by  Eqs. (\ref{bc3}).
The same steady condition gives the relationship between 
$\nabla_{\bf x} \bf U$,  $\Lambda$ and $R_T$, which represents the equation of the kinetic 
energy in the case of steady homogeneous turbulence
\bea
\frac{S} {\Lambda^2} = \frac{15}{R_T} 
\label{steady E}
\eea
where $S$ is related to $\nabla_{\bf x} \bf U$
\bea
S = \frac{\partial U_i}{\partial x_k}  
\left( \frac{\partial U_k}{\partial x_i}   
+  \frac{\partial U_i}{\partial x_k}  \right) 
\eea
The tensor  $R_{i k}$ is  then calculated with Eq. (\ref{R}) and (\ref{R0}), by means of $f$.
 
\cite{deDivitiis_3} studies some of the properties of Eqs. (\ref{I2 steady})-(\ref{bc3}), 
and shows that their solutions behave like $f -1 \approx r$ in a given interval of $r$. Accordingly, the statistical moments of velocity difference are 
$\langle \Delta u^n \rangle \approx r^{n/2}$ for $n <5$, 
and the energy spectrum $E(\kappa) \approx k^{-2}$ in the inertial subrange.

 Equation (\ref{3}) was obtained for an arbitrary flow,
 whereas the other equations were derived  under the assumption 
of steady homogeneous turbulence with a given average velocity gradient.
The successive sections show that these equations can be applied also in the case of non-homogeneous turbulence where $\lambda_T$ and $u$  vary with the space coordinates. 
Thus, the method is here applied to the turbulent region of the fully developed boundary layer, where the several quantities vary with the wall normal coordinate.

\bigskip

\section{\bf Analysis \label{Analysis}}

An evolution equation for the velocity correlation 
is determined, in the case of non-homogeneous turbulence with a nonzero
average velocity gradient.

The fluid velocity, measured in the reference frame $\Re$, is
${\bf v} ={\bf U} + {\bf u}$, where ${\bf U} \equiv (U_x, U_y, U_z)$ and 
${\bf u} \equiv (u_x, u_y, u_z)$ are, average and fluctuating velocity, respectively.
The velocity correlation tensor is $R_{i j} = \langle u_i u'_j \rangle$, where 
 $u_i$ and $u'_j$ are the velocity components of $\bf u$ calculated at $\bf x$ 
and ${\bf x'} = {\bf x} + {\bf r}$, and $\bf r$ is the separation distance.
As the analysis is finalized to the description of the turbulent region of the developed boundary layer, and since there the effects of the spatial variations of 
$\nabla_{\bf x} {\bf U}$ are orders of magnitude much smaller than those caused by 
$\nabla_{\bf x} {\bf U}$ (\cite{Karman30}, \cite{Nikuradse}), the velocity gradient is assumed to be a function of $\bf x$ alone,  being $\nabla_{\bf x} {\bf U} = \nabla_{\bf x} {\bf U}'$.

In order to determine the evolution equation of $R_{i j}$, the Navier-Stokes equations are written for the fluctuating velocity in the points ${\bf x}$ and ${\bf x}'$.
The evolution equation of $R_{i j}$ is determined by multiplying first and second equation by $u'_j$ and $u_i$, respectively, summing the so obtained equations, and calculating the average on the statistical ensemble (\cite{Karman38}, \cite{Batchelor53})
\bea
\ds
\frac{\partial R_{i j}}{\partial t}=
T_{i j} + P_{i j} 
+ 2 \nu \nabla^2 R_{i j} 
- \frac{\partial U_i}{\partial x_k} R_{k j}
- \frac{\partial U_j}{\partial x_k} R_{i k}  
+ \frac{\partial R_{i j} }{\partial r_k} 
(U_k - U_k') + \Gamma_{i j}
\label{cc1 A}
\label{cc1}
\eea
being
\bea
\begin{array}{l@{\hspace{+0.2cm}}l}
\ds T_{i j} ({\bf x}, {\bf r}) =\frac{\partial }{\partial r_k} 
\left\langle u_i  u_j' (u_k - u_k')  \right\rangle, \ \ \ \
\ds P_{i j} ({\bf x}, {\bf r}) = \frac{1}{\rho} 
 \frac{\partial \langle p u'_j \rangle}{\partial r_i} 
\end{array}
\eea
and $p$ is the fluctuating pressure.
The quantities of Eq. (\ref{cc1}), which in turn depend on $\bf r$, due to non-homogeneity depend also on $\bf x$, and Eq. (\ref{cc1}) incudes an additional term with respect to the homogeneous turbulence, represented by $\Gamma_{i j}({\bf x}, {\bf r})$ (\cite{Oberlack})
\bea
\begin{array}{l@{\hspace{+0.2cm}}l}
\ds \Gamma_{i j} ({\bf x}, {\bf r})= 
-\frac{1}{\rho} \frac{\partial \langle p u_j' \rangle }{\partial x_i} 
-\frac{\partial}{\partial x_k } \langle u_i u_k u_j' \rangle
 - \frac{1}{\rho} \frac{\partial \langle p' u_i \rangle }{\partial r_j} 
\ds + \nu  \frac{\partial^2 R_{i j} }{\partial x_k \partial x_k } \\\\
\ds -2 \nu \frac{\partial^2 R_{i j} }{\partial r_k \partial x_k }  
\end{array}
\eea
which provides the non-homogeneity of the different terms of correlation.

Making the trace of Eq. (\ref{cc1 A}), we obtain the following scalar equation
\bea
\ds
\frac{\partial R}{\partial t}=
\frac{1}{2} H 
+ 2 \nu \nabla^2 R 
- \frac{\partial U_i}{\partial x_k} R_{i k}^S 
+ \frac{\partial R }{\partial r_k} 
(U_k - U_k') + \Gamma
\label{cc2}
\eea
being $R_{i k}^S$ is the symmetric part of $R_{i k}$ and
\bea
\ds R = \frac{R_{i i}}{2} ,  \ \ \ \Gamma = \frac{\Gamma_{i i}}{2} 
\eea
$R({\bf x}, 0)$ gives the turbulent kinetic energy, $H$ $\equiv$ $T_{i i}$ provides the mechanism of energy cascade (\cite{Karman38},  \cite{Batchelor53}), 
$P_{i i} \equiv 0$ expresses the fluid incompressibility whereas $\Gamma$ 
arises from the non-homogeneity of the flow.

Now, a scalar equation for describing the
main properties of the velocity correlation is determined.
To this end,  $R_{i j}$,  $H$ and $\bf U$ are decomposed into an even function of 
$r \equiv \vert {\bf r} \vert$ (here called spherical part),  plus the remaining term:
\bea
\begin{array}{l@{\hspace{+0.2cm}}l}
R_{i j} ({\bf x}, {\bf r})= \hat{R}_{i j}({\bf x}, r) + \Delta R_{i j} ({\bf x}, {\bf r}) \\\\
H ({\bf x}, {\bf r}) = \hat{H}({\bf x}, r) + \Delta H ({\bf x}, {\bf r}) \\\\
{\bf U}' -{\bf U} = \hat{\bf U}({\bf x}, r) + \Delta {\bf U} ({\bf x}, {\bf r}) 
\end{array}
\label{dec}
\eea 
where $\hat{F}$ is the spherical part of the generic quantity $F$, defined as
\bea
\begin{array}{l@{\hspace{+0.99cm}}l}
\ds \hat{F}({\bf x}, r) =  \frac{1}{6} \left( F({\bf x}, r, 0, 0) + F({\bf x}, 0, r, 0)+ F({\bf x}, 0, 0, r) \right) \\\\ 
\ds + \frac{1}{6} \left( F({\bf x}, -r, 0, 0) + F({\bf x}, 0, -r, 0)+ F({\bf x}, 0, 0, -r) \right)
\end{array}
\label{isotrp}
\eea
and $\Delta R_{i j}({\bf x}, {\bf 0})$=$\Delta H({\bf x}, {\bf 0})$ = 0.
Therefore, the Fourier transform of $\hat{R}$ identifies the part of the energy spectrum depending upon $\bfkappa^2$.
As the consequence of this decomposition, $R$ and $\Delta R$ satisfy the equations 
\bea
\begin{array}{l@{\hspace{+0.2cm}}l}
\ds \frac{\partial \hat{R}}{\partial t} =
\frac{\hat{H}}{2} 
+ 2 \nu \left(  \frac{\partial^2 \hat{R}} {\partial r^2} +
\ds \frac{2}{r} \frac{\partial \hat{R}}{\partial r}  \right)  
- \hat{G}
\end{array}
\label{eq iso}
\eea
\bea
\begin{array}{l@{\hspace{+0.2cm}}l}
\ds \frac{\partial \Delta R}{\partial t}  =
\frac{ \Delta H}{2}
+ 2 \nu \nabla^2 \Delta R 
- \Delta{G}
\end{array}
\label{eq non-iso}
\eea
in which Eq. (\ref{eq non-iso}) is obtained as the difference between Eqs. 
(\ref{cc2}) and (\ref{eq iso}), and
\bea
\begin{array}{l@{\hspace{+0.2cm}}l}
\ds \hat{G} = \frac{\partial U_i} {\partial x_k} \hat{R}_{k i}^S + \hat{G}_0, \ \ \ \
\ds \Delta{G} = \frac{\partial U_i} {\partial x_k} \Delta{R^S}_{k i} 
- \Delta \left( \frac{\partial R }{\partial r_k} (U_k - U_k')\right) -\Delta \Gamma
\end{array}
\eea
and $\hat{G}_0$ represents the spherical part of $-\partial R /\partial r_k (U_k - U_k') -\Gamma $.
It is worth to remark that Eq. (\ref{eq iso}) formally coincides with the equation obtained
in \cite{deDivitiis_3},
with the difference that here, because of non-homogeneity, the quantities appearing into 
Eq. (\ref{eq iso}) depend also on $\bf x$.

The turbulence is here studied using Eq. (\ref{eq iso}) 
alone, whereas $R_{i j}({\bf x}, {\bf r})$ will be determined in function of 
$\nabla_{\bf x}{\bf U}$, by means of a proper statistical analysis 
of the two-points velocity correlation.

\bigskip

\section{\bf Pair distribution function}
 \label{distribution function}

This section analyses the non-homogeneous turbulence through the pair distribution function, taking into account that $\lambda_T$ and $\langle u_i u_j \rangle$ vary  with the spatial coordinates, whereas $\nabla_{\bf x} {\bf U}$ is considered to be an assigned quantity.
To study this, consider now the pair distribution function of the fluid velocity
\bea
F^{(2)} ({\bf v}, {\bf v}'; {\bf x}, {\bf x}') =  F^{(2)}_0 \left( {\bf v}, {\bf v}'; {\bf x}, {\bf x}' \right)  + \phi^{(2)} ({\bf v}, {\bf v}'; {\bf x}, {\bf x}')
\label{iso00}
\eea
where $F^{(2)}$ obeys to Eq. (\ref{3}), and $\phi^{(2)}$, representing the deviation from the isotropic turbulence, satisfies, at each instant, the following equations
(\cite{deDivitiis_3})
\bea
\begin{array}{l@{\hspace{+0.2cm}}l}
\ds \int_{v} \int_{v'} \phi^{(2)} du^3 du'^3= 0, \\\\
\ds \int_{v} \int_{v'} \phi^{(2)} {\bf u}  \ du^3 du'^3= 0, \\\\
\ds \int_{v} \int_{v'}  \phi^{(2)} {\bf u} \cdot {\bf u} \ du^3 du'^3= 0.
\end{array}
\label{cons1}
\eea
Equations (\ref{cons1}) state that momentum and kinetic energy associated to $F^{(2)}_0$ and $F^{(2)}$ are equal each other, respectively.
The analysis supposes also that all the dimensionless statistical moments of $F^{(2)}_0$ are constant in all the points of the fluid domain, therefore 
the functional form of $F^{(2)}_0$ is assumed to be
\bea
F^{(2)}_0 ({\bf v}, {\bf v}'; {\bf x}, {\bf x}') =  
F^{(2)}_0 \left( \frac{{\bf v} -{\bf U} ({\bf x})}{u({\bf x})}, \frac{{\bf v}'-{\bf U} ({\bf x}')}{u( {\bf x}')} \right)  
\label{iso0}
\eea
and its gradients are calculated in functions of the spatial derivatives of $\bf U$ and $u$
\bea
\frac{\partial F^{(2)}_0}{\partial x_k} = 
- \frac{\partial F^{(2)}_0}{\partial v_j} 
\left( \frac{\partial U_j}{\partial x_k} + \frac{u_j}{u} \frac{\partial u}{\partial x_k}\right) ,
 \ \ \ \
\frac{\partial F^{(2)}_0}{\partial x'_k} = 
-\frac{\partial F^{(2)}_0}{\partial v'_j} 
\left( \frac{\partial U_j}{\partial x_k} + \frac{u_j'}{u'} \frac{\partial u'}{\partial x_k'}\right) 
\label{4}
 \eea 
where, as before,  $\nabla_{\bf x} {\bf U} = \nabla_{\bf x'} {\bf U}'$.
Substituting Eqs.(\ref{iso00}) and (\ref{4}) into Eq. (\ref{3}), we obtain the following relationship between $F^{(2)}_0$ and $\phi^{(2)}$
\bea
\begin{array}{l@{\hspace{+0.2cm}}l}
\ds \lambda \phi^{(2)} = 
 -J_D   
\ds - \left(  \frac{\partial F^{(2)}_0 }{\partial t} 
+ \frac{\partial \phi^{(2)} }{\partial t} 
+  \frac{\partial \phi^{(2)} }{\partial x_p} v_p 
+  \frac{\partial \phi^{(2)} }{\partial x_p'} v_p'
  \right) \\\\
 \hspace{+15 mm} 
\ds +\left(
\frac{\partial F^{(2)}_0}{\partial v_j} \frac{v_p u_j}{u} \frac{\partial u}{\partial x_k}\right)
+ \left(\frac{\partial F^{(2)}_0}{\partial v'_j}  \frac{ v'_p u'_j}{u'} \frac{\partial u'}{\partial x'_k} \right)
 \\\\
 \hspace{+15 mm} 
\ds +\left(
\frac{\partial F^{(2)}_0}{\partial v_j}  v_p 
+ \frac{\partial F^{(2)}_0}{\partial v'_j}  v'_p 
\right)  \frac{\partial U_j}{\partial x_p} 
\end{array}
\label{fi_2}
\eea
As the last term of Eq. (\ref{fi_2})
identically satisfies Eqs. (\ref{cons1}), these latter are written in the form  
\bea
\begin{array}{l@{\hspace{+0.2cm}}l}
\ds \int_v \hspace{-1.0mm} \int_{v'} \hspace{-2.0mm}
\left[\begin{array}{c}
 \ds 1 \\\
 \ds {\bf u}  \\\
 \ds {\bf u} \cdot {\bf u} 
\end{array}\right]  
\ds   (  \frac{\partial F^{(2)} }{\partial t} 
+ \frac{\partial \phi^{(2)} }{\partial x_p} v_p
+  \frac{\partial \phi^{(2)} }{\partial x_p'} v_p'
 - \frac{\partial F^{(2)}_0}{\partial v_j} \frac{v_p u_j}{u} 
   \frac{\partial u}{\partial x_k} \\\\
\ds - \frac{\partial F^{(2)}_0}{\partial v'_j}  \frac{ v'_p u'_j}{u'} \frac{\partial u'}{\partial x'_k}
  +  J_D   )  \ du^3 du'^3 \equiv 0
\end{array}
\label{c0}
\eea
A sufficient condition for satisfying these five equations is that the integrand of
 Eq. (\ref{c0}) vanishes 
\bea
\frac{\partial F^{(2)} }{\partial t} 
+  \frac{\partial \phi^{(2)} }{\partial x_p} v_p
+  \frac{\partial \phi^{(2)} }{\partial x_p'} v_p'
- \frac{\partial F^{(2)}_0}{\partial v_j} \frac{v_p u_j}{u} \frac{\partial u}{\partial x_k}
- \frac{\partial F^{(2)}_0}{\partial v'_j}  \frac{ v'_p u'_j}{u'} \frac{\partial u'}{\partial x'_k}
+  J_D   \equiv 0
\label{sc}
\eea
Assuming that Eq. (\ref{sc}) is true, and taking into account Eq. (\ref{fi_2}),
$F^{(2)}$ is given by Eq. (\ref{F2}) which does not depend on the gradient of $u$ and $u'$, 
as in the case of the homogeneous turbulence.
Really, $F^{(2)}$ changes starting from an arbitrary initial condition, therefore
Eq. (\ref{F2}) represents an approximation which can be considered to be valid far from the initial condition. 
As the consequence, also the spherical part of the correlation tensor,  $\hat{R}_{i j}$, here obtained
\bea
\hat{R}_{k i} = \hat{R}^S_{i k} = \frac{u^2}{3} \left(3 f + \frac{\partial f}{\partial r} r \right) \left( \delta_{k i} 
-\frac{1}{\lambda} \left( \frac{\partial U_k}{\partial x_i}   
+  \frac{\partial U_i}{\partial x_k}    \right)  \right) 
\label{R2}
\eea
coincides with the expression given in \cite{deDivitiis_3}.
Thus, substituting Eq. (\ref{R2}) and into Eq. (\ref{eq iso}) and following the analytical procedure of \cite{deDivitiis_3}, we found that  $f$ obeys to Eq. (\ref{I2 steady}) also in the present case, and $\Lambda$ is related to  $\nabla_{\bf x} {\bf U}$ and $R_T$ through Eq. (\ref{steady E}). The difference with respect to \cite{deDivitiis_3} is that, here the turbulence is non-homogeneous, thus $R_T = R_T({\bf x})$, and $f$ also depends on $\bf x$, being 
$f = f({\bf x}, r)$.

\bigskip

\section{\bf Analysis of the boundary layer}

Here, the steady turbulent boundary layer with a moderate pressure gradient is analyzed,
 assuming  that the flow is fully developed along the streamwise direction.
To this purpose, return to Fig. \ref{figura_1}, 
and consider only the developed region of the flow.
In the figure, $\Re$ is the wall frame of reference, $x$ and $y$ are, respectively, the streamwise direction and the coordinate normal to the wall, whereas $z$ is the spanwise coordinate.

In the developed region, $u$,  $\langle u_x u_y \rangle$ and $\lambda_T$ 
change with $y$, and  in the laminar sublayer $U_x$ and $u$ are both about proportional to $y$.
In LL, BL and at the beginning of TR, the correlation scale of velocity is proportional to the distance from the wall, and vanishes for $y=0$, being
\bea
\ds \lambda_T = \left( \frac{\partial \lambda_T}{\partial y} \right)_0 y + ... 
\label{lambda e}
\eea 
where $\left( {\partial \lambda_T}/{\partial y} \right)_0 = O(1)$, whereas
the velocity fluctuations follow the Navier-Stokes equations and satisfy the wall
 boundary conditions
\bea
\begin{array}{l@{\hspace{+0.2cm}}l}
\ds u_x = \frac{\partial u_x}{\partial y}(0) y + ..., \ \ \
\ds u_y = \frac{1}{2} \frac{\partial^2 u_y}{\partial y^2}(0) y^2 + ..., \ \ \
\ds u_z = \frac{\partial u_z}{\partial y}(0) y + ... 
\end{array}
\label{bc}
\eea

In case of fully developed flow along $x$, that is, parallel flow assumption 
($\partial / \partial y$ $>>>$ $\partial / \partial x$), the continuity and momentum equations
of the mean flow are (\cite{Schlichting})
\bea
\begin{array}{l@{\hspace{+0.2cm}}l}
\ds \frac{\partial U_y}{\partial y} =0,
\end{array}
\label{continuity 2}
\eea
\bea
\begin{array}{l@{\hspace{+0.2cm}}l}
\ds  \frac{\partial}{\partial y} \langle u_x u_y \rangle  
\ds + \frac{1}{\rho} \frac{\partial P}{\partial x} 
\ds - \nu \frac{\partial^2 U_x}{\partial y^2}=0, \\\\
\ds  \frac{\partial}{\partial y} \langle u_y^2 \rangle  
\ds + \frac{1}{\rho} \frac{\partial P}{\partial y} =0
\end{array}
\label{momentum 2}
\eea
where $P$ is the average pressure, whereas the equation of the turbulent kinetic energy reads as
\bea
\ds  \frac{\partial}{\partial y}  \left\langle u_y  \left( \frac{p}{\rho} 
+ \frac{u_j u_j}{2}\right)  \right\rangle
\ds + \left\langle u_x u_y \right\rangle  \frac{\partial U_x}{\partial y}
\ds  + \nu \left\langle  \frac{\partial u_j}{\partial x_i} \frac{\partial u_j}{\partial x_i} \right\rangle  =0
\label{energy 2}
\eea
The non-homogeneity is responsible for the first term
of Eq. (\ref{energy 2}), whereas second and third terms represent, respectively, 
the energy production due to the average motion, and the dissipation.

\bigskip

Because of the  parallel flow assumption, in all these equations $U_x$, $\langle u_i u_j\rangle$ are functions of $y$ alone. 
From Eq. (\ref{continuity 2}) and taking into account the boundary conditions, $U_y \equiv 0$,
whereas Eqs. (\ref{momentum 2}) give the following first integrals
\bea
\begin{array}{l@{\hspace{+0.2cm}}l}
\ds \frac{P(x, y)}{\rho} = F x + H - \langle u_y^2 \rangle
\end{array}
\label{y}
\eea
\bea
\begin{array}{l@{\hspace{+0.2cm}}l}
\ds \langle u_x u_y \rangle -\nu \frac{d U_x}{d y}=
- F y -U_T^2
\end{array}
\label{x}
\eea
being  
$F= 1/\rho \ \partial P/ \partial x$
and $H$ is a proper constant proportional to the average pressure at $x=0$.
Into Eq. (\ref{x}), the boundary layer approximation and the hypothesis 
of moderate average pressure gradient along $x$ provide that 
$ U_T^2 >> \vert F y \vert$ in all the regions, therefore introducing the dimensionless variables $U^+ = U_x/U_T$ and $y^+ = y U_T /\nu$, Eq. (\ref{x}) reads as 
\bea
\begin{array}{l@{\hspace{+0.2cm}}l}
\ds  \frac{d U^+}{d y^+}= \frac{1}{1+R_T}
\end{array}
\label{x1}
\eea
Equation (\ref{x1}) is assumed to describe the average flow with 
moderate pressure gradient in the three regions of the boundary layers,
when the parallel flow hypothesis is verified. 

\bigskip

\subsection{\bf The Turbulent Region}

This section studies the distribution of the different variables in 
the turbulent region TR, by means of the analysis seen in the sections \ref{Analysis} and \ref{distribution function}.

First, observe that,  in case of fully developed parallel flow,
the effects of non-homogeneity in TR are much smaller than those related
to energy production and dissipation (\cite{Tennekes}), thus the first term of Eq. (\ref{energy 2}) is
here neglected with respect to the other ones, and Eq. (\ref{steady E}) is recovered.
This approximation, in agreement with the analysis seen in sect. \ref{distribution function}, states that the kinetic energy production is balanced only by the dissipation in TR, and allows to express the several variables in terms
of the local value of $R_T$ (or $dU_x/dy$).

In order to calculate $u^+ = u/U_T$,  $dU_x/dy$ is eliminated between 
Eqs. (\ref{x}) and (\ref{steady E}), where $S = (dU_x/dy)^2$
\bea
\ds u^+ = \frac{R_T^{3/4}}{15^{1/4} \sqrt{1+R_T}} 
\label{u}
\eea
Being $R_T = u^+ \lambda_T^+$, also $\lambda_T^+$ is in terms of $R_T$
\bea
\ds \lambda_T^+ = (15 R_T)^{1/4} \sqrt{1+R_T}
\label{lambda}
\eea
As these equations arise from Eq. (\ref{steady E}) which holds in the turbulent
region,  Eqs.(\ref{u}) and (\ref{lambda}) describe the variations of $u$ and $\lambda_T$
only in TR.

From the comparison between Eqs. (\ref{x1}) and (\ref{vk alt}), $\left( {d R_T}/{d y^+}\right)_e$ identifies the von K\'arm\'an constant, here expressed taking into account that 
$R_T \equiv \lambda_T^+ u^+$
\bea
k \equiv \left( \frac{d R_T}{d y^+}\right)_e = 
\left( \frac{d u^+}{d y^+}\right)_e \lambda_{T e}^+ +
\left( \frac{d \lambda_{T}^+}{d y^+}\right)_e u_e^+
\label{vkc}
\eea
where the subscript $e$ denotes the values calculated at the edge of TR.
Substituting Eqs. (\ref{u}) and (\ref{lambda}) into Eq. (\ref{vkc}), we obtain the von K\'arm\'an constant in function of the variables at $y^+_e$
\bea
\ds k = \frac{4}{15^{1/4}} \left( \frac{d \lambda_{T}^+}{d y^+}\right)_e
\ds \frac{\sqrt{1+R_{T e}}}{1+ 3 R_{T e}}  R_{T e}^{3/4}
\label{vkc1}
\eea
It is worth to remark that this estimation of $k$ requires the knowledge of $R_{T e}$ and
of $(d\lambda_T^+/dy^+)_e$, whereas does not
need the assumption that $U^+$ is represented by a logarithmic profile.

Next, the dimensionless Prandtl's mixing scale $l_p^+$ is calculated from the definition
of  Prandtl's mixing length $l_p$
\bea
- \langle u_x u_y \rangle =  \left| \frac{\partial U_x}{\partial y}  \right|
 \frac{\partial U_x}{\partial y} \ l_p^2
\label{Prandtl_d}
\eea
and Eqs. (\ref{u_x}) and (\ref{steady E}). This is
\bea
l_p^+ = \sqrt{R_T (1 + R_T)} 
\label{Prandtl}
\eea
being $l_p^+ \simeq R_T$ for $R_T >> 1$,  and its  derivative calculated at $y^+_e$ is also
expressed in terms of $R_{T e}$
\bea
\ds \left( \frac{d l^+_p}{d y^+} \right)_e = \frac{1+2 R_{T e}}{2 \sqrt{R_{T e}(1+R_{T e})}} \ k 
\label{Prandtl_y}
\eea
In view of Eq. (\ref{vkc1}), this derivative can be also expressed in function of $d \lambda_T^+ /d y^ì$
\bea
\ds \left( \frac{d l^+_p}{d y^+} \right)_e = 2 \ \frac{1+2 R_{T e}}{1+ 3 R_{T e}} \ 
\left( \frac{ R_{T e} }{15}\right)^{1/4}  \left( \frac{d \lambda_{T}^+}{d y^+}\right)_e
\ds 
\label{Prandtl_y1}
\eea
This expression gives the link between the variations of the Taylor scale and of the Prandtl's length at the border of the turbulent region.

Following  Eqs. (\ref{u}) and (\ref{lambda}), $u^+$ and $\lambda_T^+$ are functions of $y^+$ through the local value of $R_T$ (or of $dU^+/dy^+$), therefore, the distribution of such quantities along $y^+$ require the knowledge of the function $R_T=R_T(y^+)$. 
This latter can be expressed as 
\bea
\ds R_T(y^+) = R_{T e} + \left( \frac{d R_T}{d y^+}\right)_e 
\left( y^+ - y^+_e \right) + O \left( y^+ - y^+_e \right)^2
\label{R_T}
\eea
Now, in a range of $y^+$ where $O \left( y^+ - y^+_e \right)^2$ is negligible with respect to the other terms, $d U^+/d y^+$ is 
\bea
\ds \frac{d U^+}{d y^+}= \ds \frac{1}{\ds 1 +  R_{T e} + \left( \frac{d R_T}{d y^+}\right)_e 
\ds \left( y^+ - y^+_e \right)}
\label{vd}
\eea
$U^+$ exhibits there logarithmic law, obtained integrating Eq. (\ref{vd}) 
from $y^+_e$ to $y^+$
\bea
U^+ = \frac{1}{k} \ln \left( \frac{1+R_{T e} + k (y^+-y^+_e)}{1+R_{T e}} \right) + U^+_e
\label{U turbo}
\eea
being $U^+_e=U^+(y^+_e)$.
This law is defined as soon as the parameters $y^+_e$, $k$ and $U^+_e$
are known. Equation (\ref{U turbo}) differs from the classical expression (\ref{wall law}) 
and formally tends to Eq. (\ref{wall law}) when $y^+ \rightarrow \infty$.
Therefore, Eq. (\ref{U turbo}) can give values of $U^+$ sizably different from (\ref{wall law}) for small $y^+$.
The comparison between these equations, for $y^+ \rightarrow \infty$ identifies 
$C$ in terms of the variables at $y_e$
\bea
C = U^+_e + \frac{1}{k} \ln \left( \frac{k}{1+R_{T e}}\right) 
\label{C}
\eea

\bigskip

As far as the Reynolds stress is concerned, it is expressed in function of $R_T$ 
through Eqs. (\ref{x}) and (\ref{x1}) 
\bea
\begin{array}{l@{\hspace{+0.2cm}}l}
\ds \langle u v \rangle^+ \equiv \frac{\langle u_x u_y \rangle}{U_T^2} = - \frac{R_T}{1+R_T}
\end{array}
\label{uv}
\eea
Observe that, the wall boundary conditions (\ref{bc})
state that $\langle u v \rangle^+ \approx {y^+}^3$ near the wall,
whereas Eq. (\ref{uv}) gives  $\langle u v \rangle^+ \approx R_T \approx {y^+}^2$.
This disagreement is due to the fact that the expression of $\langle u v \rangle^+$
has not been derived from the correlation equation with $r=0$, but arises from
Eq. (\ref{x}) eliminating $dU_x/dy$. This implies that Eq. (\ref{uv}) holds only in TR and BL, whereas in the laminar sub-layer a proper matching condition must be applied.

\bigskip

\subsection{\bf  Matching Turbulent region - buffer layer}

With reference to Fig. \ref{figura_1}, the domains LL and BL constitute SL,
a zone between wall and turbulent region.
There, due to the presence of the wall, the analysis of sections \ref{distribution function}
and \ref{Analysis} can not be applied.
Therefore, the mean variables in SL are expressed in function of $y^+$,
taking into account the boundary conditions  (\ref{bc}) and that  $\lambda_T$
follows Eq. (\ref{lambda e}), where it is assumed 
$(d \lambda_T^+/dy^+)_e = (d \lambda_T^+/dy^+)_0$.
As the consequence, $R_T$, $u^+$
 and $\lambda_T^+$ are supposed to vary in SL according to
\bea
\begin{array}{l@{\hspace{+0.2cm}}l}
\ds  R_T = u^+ \lambda_T^+, \\\\
\ds u^+ = {u^+_e}  \  \frac{y^+ +C_u {y^+}^2}{y^+_e +C_u {y^+_e}^2}, \  \  \ (0 < y^+ \le y^+_e) \\\\
\ds \lambda_T^+ = \lambda_{T e}^+ \frac{y^+}{y^+_e}, \  \ \ (0 < y^+ \le y^+_e),
\end{array}
\label{ss}
\eea
where $C_u$ is a constant given by
\bea
\begin{array}{l@{\hspace{+0.2cm}}l}
\ds C_u = \frac{1}{y^+_e} \frac{2 R_{T e}- k  y^+_e}{k y^+_e -3 R_{T e}} 
\end{array}
\label{ssc}
\eea
Equations (\ref{ss}) and (\ref{ssc}) provide the matching condition between SL and TR.
Specifically, Eqs. (\ref{ss}) state that $u^+$, $R_T$ and
$\lambda_T^+$ are continuous functions for $y^+ = y^+_e$, whereas 
Eq. (\ref{ssc}) gives there the continuity of their derivatives.
The average velocity is then calculated by quadrature, substituting the expression of $R_T$ given by Eqs. (\ref{ss}), into Eq. (\ref{x1}),
integrating this latter from 0 to $y^+$, with $U^+(0) = 0$
\bea
U^+(y^+) = \int_0^{y^+} \frac{1}{1+R_T(\eta^+)} d \eta^+
\label{U ss}
\eea
where $U^+(y^+) \simeq y^+$ for small $y^+$.
The matching between SL and TR gives the value of $U^+_e$.
\bea
U^+_e = \int_0^{y^+_e} \frac{1}{1+R_T(\eta^+)} d \eta^+
\eea
Thus, Eq. (\ref{U ss}) and (\ref{U turbo}) establish that both $U^+$ and $d U^+ / d y^+$ are continuous
for $y^+ = y^+_e$

\bigskip

\subsection{\bf  Matching buffer layer - laminar sub-layer}

As previously seen, the expression  (\ref{uv}) of the Reynolds stress,  valid
in  BL  and TR, can not be applied in the laminar region.
Since $\langle u v \rangle^+ \approx {y^+}^3$ near the wall,
the Reynolds stress in LL is approximated by  
\bea
\ds \langle u v \rangle^+ = \langle u v \rangle^+_*
\ds \frac{ {y^+}^3 +C_{uv} {y^+}^4 } {{y^+_*}^3 +C_{uv} {y^+_*}^4}, \ \ \ (0 < y^+< y^+_*)
\label{ss uv}
\eea
being $C_{uv}$ a constant 
\bea
\ds C_{uv} = \frac{1}{y^+_*} \frac{3- (d \ln \langle uv \rangle^+/ d y^+)_* 
y^+_*} {(d \ln \langle uv \rangle^+/ d y^+)_* y^+_* -4}
\label{ss uv c}
\eea
Equations (\ref{ss uv}) and (\ref{ss uv c}) establish that the Reynolds stress and
its derivative are continuous for $ y^+ = y^+_*$, where the subscript $*$ indicates 
the value calculated at $ y^+_*$. 
 This latter, obtained as the inflection point of  
$\langle u v \rangle^+$ (i.e. $d^2 \langle u v \rangle^+/{dy^+}^2 =0$) according to Eqs. (\ref{uv}) and (\ref{ss}), gives the dimensionless tickness of the laminar sub-layer.

\bigskip

\section{\bf Identification of the velocity law free parameters}

The definition of the velocity law, requires the knowledge of the 
parameters which appear into Eqs. (\ref{vkc1}) and (\ref{U turbo}).
In particular, the determination of $k$ needs the values of $R_{T e}$ and 
$(d \lambda_T^+/d y^+)_e$.
To identify these latter, we will proceed as follow.

First, observe that $\langle u_i u_j \rangle$ is a symmetric tensor which can be obtained from the diagonal tensor
\bea
\langle {\bf u}_0 {\bf u}_0 \rangle =
\left[\begin{array}{ccc}
 \ds \langle u_0^2 \rangle  & 0 & 0  \\\\
 \ds 0 & \langle v_0^2 \rangle & 0   \\\\
 \ds 0 & 0 & \langle w_0^2 \rangle  
\end{array}\right]  
\eea
through an opportune rotation around $y \equiv y_0$, being  $\langle u_0^2 \rangle$,   $\langle v_0^2 \rangle$ and  $\langle w_0^2 \rangle$ the velocity components standard deviations in the canonical frame, and $\vartheta$ is the angle of this rotation.
Therefore, $u$ and $\langle u_x u_y \rangle$ are in terms of the elements of 
$\langle {\bf u}_0 {\bf u}_0 \rangle$
\bea
\begin{array}{l@{\hspace{+0.2cm}}l}
\ds u^2 = \frac{1}{3} (\langle u_0^2 \rangle + \langle v_0^2 \rangle +\langle w_0^2 \rangle) \\\\
\ds \langle u_x u_y \rangle = (\langle u_0^2 \rangle - \langle v_0^2 \rangle) \frac{\sin 2 \vartheta}{2} 
\end{array}
\label{cr}
\eea
Because of the parallel flow assumption and taking into account Eqs. (\ref{u_x}) and (\ref{cr}),  $u$ and $\langle u_x u_y \rangle$ are related each other in such a way that 
\bea
u^2 \ge \vert \langle u_x u_y \rangle \vert
\eea
Hence, Eq. (\ref{steady E}) implies that
\bea
R_T \ge 15  \ \ \ {\mbox{or}} \ \ \ \frac{d U_x}{d y} \lambda_T \ge u
\eea
in the turbulent region.
This limitation identifies the minimum value of $R_T$ in TR, which is assumed to be 
\bea
R_{T e} = 15.
\eea

To estimate $(d \lambda_T^+/d y^+)_e$, consider first the spanwise correlation function of the streamwise velocity components 
(that is $R_{1 1} (r_z)$ = $\langle u_x(x, y, z) u_x(x, y, z+r_z) \rangle$).
This can be calculated with Eq. (\ref{R}), once $f$ is known through 
Eqs. (\ref{I2 steady}).
Since $\partial U / \partial y$ leads to the development of coherent structures in the fluid, similar to streaks and caused by the stretching of the vortex lines along
the streamwise direction  (\cite{Kim}), we expect that $R_{1 1} (r_z)$  
intersects the horizontal axis and remains negative for $r_z \rightarrow \infty$.
This implies a wide distribution of spacings between the different streaky
structures, whose mean value depends on $R_T$ (\cite{Kim}). 
Accordingly, there exists a minimum spanwise distance $r_{z}^*$
such that, for $r_{z} \in \left[0, r_{z}^* \right]$, $R_{1 1} (r_z)$ gives the necessary informations to describe the statistical properties of the correlation  
(\cite{Ventsel}).
  \begin{figure}[t]
	\centering
         \includegraphics[width=0.60\textwidth]{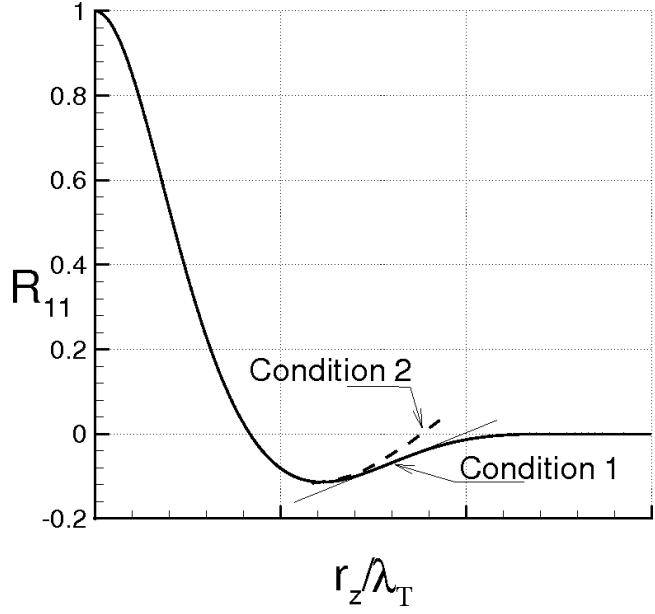}
\caption{Schematic of the spanwise correlation function of $u_x$ at the edge of the turbulent region and definition of Eqs. (\ref{condizio 1}) and (\ref{condizio 2}).}
\label{figura_2}
\end{figure}
According to the theory (\cite{Ventsel}), if $R_{1 1}$ monotonically tends to zero as
$r_z \rightarrow \infty$ and $R_{1 1} < 0$,  $r_{z}^*$ can be estimated as the distance at which $R_{1 1}$ is negative and exhibits an inflection point (second inflection point of the curve, see  Fig. \ref{figura_2})
\bea
\ds R_{1 1}(r_{z}^*) < 0, \ \ \ \  \frac{\partial^2 R_{1 1}}{ \partial r_z^2} (r_{z}^*)=0,
\ \ \ \mbox{Condition 1}
\label{condizio 1}
\eea
Alternatively, $r_{z}^*$ can be estimated as the intersection between the osculating parabola in the point $\bar{r}_{z}$  where $R_{1 1}$ = min, and the horizontal axis, that is
\bea
R_{1 1}(\bar{r}_{z}) + 
\frac{1}{2} \frac{\partial^2 R_{1 1}}{\partial r_z^2}(\bar{r}_{z}) (r_{z}^*- \bar{r}_{z})^2=0,  \ \ \ \mbox{Condition 2}
\label{condizio 2}
\eea
Since $R_{1 1} (r_z)$ is expressed in function of $f$ through Eq. (\ref{R}), this is
related to the correlation functions associated to the other directions, thus
$r_{z}^*$ is representative also for the other coordinates $x$ and $y$.
As the result, the distance from the wall of a point of TR must be always greater than $r_{z}^*$ (i.e. $r_{z}^*< y$).
Hence, it is reasonable to assume that, at the edge of the turbulent domain
\bea
\begin{array}{l@{\hspace{+0.2cm}}l}
\ds y_e = r_{z}^* \\\\
 \ds  R_{T e}=15
\end{array}
\eea


\bigskip

In order to calculate $R_{1 1}$, $f$ is first obtained by solving 
Eqs. (\ref{I2 steady})-(\ref{bc3}) which correspond
to the following Cauchy's initial condition problem  (\cite{deDivitiis_3})
\bea
\begin{array}{l@{\hspace{+0.cm}}l}
\ds  \frac{d f}{d \hat{r}} =  F \\\\
\ds \frac{d F}{d\hat{r}} = - \frac{5 f  \hat{r}}{\sqrt{2(1-f)}} -
\left( \frac{1}{2} \sqrt{\frac{1-f}{2}} R_T + \frac{4}{\hat{r}} \right) F \\\\
\ds f(0) = 1, \ F(0) = 0
\end{array} 
\label{vk-h2}  
\label{ic}
\eea
Hence, $R_{1 1} (r_z)$ is obtained with Eq. (\ref{R}), and $r_{z}^*$ is calculated  for both the conditions (\ref{condizio 1}) and (\ref{condizio 2}). 
As the result, $y^+_e$ is given by 
\bea
\frac{y_e}{\lambda_{T e}} = \frac{r_{z}^*}{\lambda_{T e}}
\ \ \ \mbox{where} \  R_T = R_{T e} \equiv 15
\eea
and $(d \lambda_T^+/d y^+)_e$ is determined according to Eqs. (\ref{ss})
\bea
\ds \left(  \frac{d \lambda_T^+}{d y^+}\right)_e = \frac{\lambda_{T e}^+}{y_e^+}
\eea

\bigskip

\section{\bf Results and Discussion}

\begin{figure}[t]
 \hspace{-7.mm}  \includegraphics[width=0.77\textwidth]{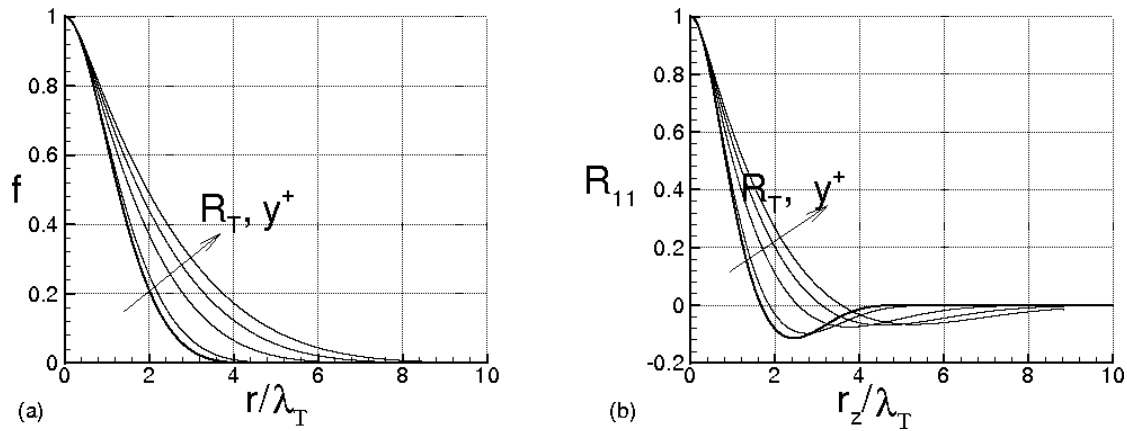}
\caption{(a) Longitudinal correlation function associated to $\hat{R}_{i j}$. (b) spanwise correlation function  of $u_x$, for different Taylor scale Reynolds number.
 The bold lines are calculated for $R_T = 15$.}
\label{figura_3}
\end{figure}

As the calculation of $k$ and $d \lambda_T^+/dy^+$ needs the knowledge
of the statistical properties of the velocity correlation,
$R_{1 1}$ is calculated for different values of $R_T$. 
To this purpose, $f$ was first determined by solving numerically Eqs. (\ref{vk-h2}),
by means of the fourth-order Runge-Kutta method. 
The calculation was carried out for $R_T$ = 15, 20, 40, 60 and 80, where these Reynolds numbers correspond to several distances from the wall in the turbulent region.
To obtain a good accuracy of the solutions, the step of integration is chosen to be  equal to 1/40 of the estimated Kolmogorov scale $\l_K$, where  
$\l_K/\lambda_T = 1/15^{1/4}/\sqrt{R_T}$ (\cite{Batchelor53}).
The results are given in Fig. \ref{figura_3} (a) which shows $f$ in terms of $r$, where the bold line represents the correlation function at the edge of TR 
($R_T = 15$). This latter exhibits the ratio (integral scale)/(Taylor scale) quite similar to that of a gaussian centered in the origin, whereas in the other cases, this ratio increases with $R_T$, in agreement with the analysis of \cite{deDivitiis_3}.
A more detailed analysis about the changing of $f$ and of the corresponding energy spectrum $E(\kappa)$ with $R_T$ is reported in \cite{deDivitiis_3}.

The spanwise correlation function $R_{1 1} (r_z)$ is then calculated with
Eq. (\ref{R}), and is represented in Fig. \ref{figura_3} (b) for the same values of $R_T$.
From these data, the edge of TR, $y^+_e \equiv r_{z}^*$, is calculated for both the conditions 1 and 2.

\begin{figure}[t]
	\centering
         \includegraphics[width=0.60\textwidth]{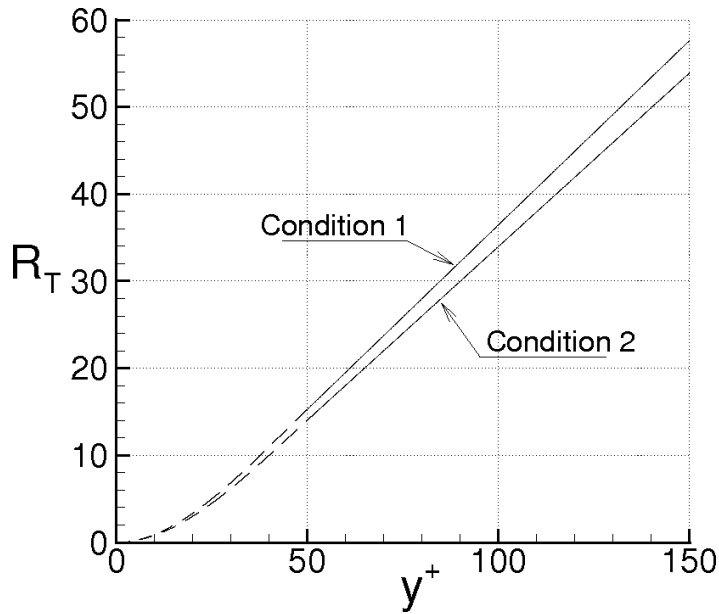}
\caption{Taylor scale Reynolds number in terms of $y^+$: SL dashed lines, 
TR continuous lines}
\label{figura_4}
\end{figure}

\begin{figure}[t]
	\centering
         \includegraphics[width=0.60\textwidth]{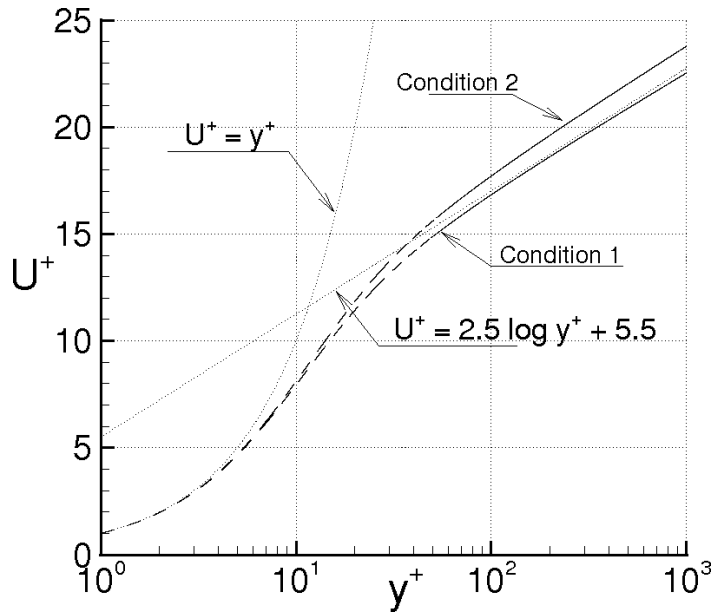}
\caption{Dimensionless average velocity profile: SL dashed lines, 
TR continuous lines}
\label{figura_5}
\end{figure}

\begin{figure}[t]
	\centering
 \hspace{-20.mm}        \includegraphics[width=0.88\textwidth]{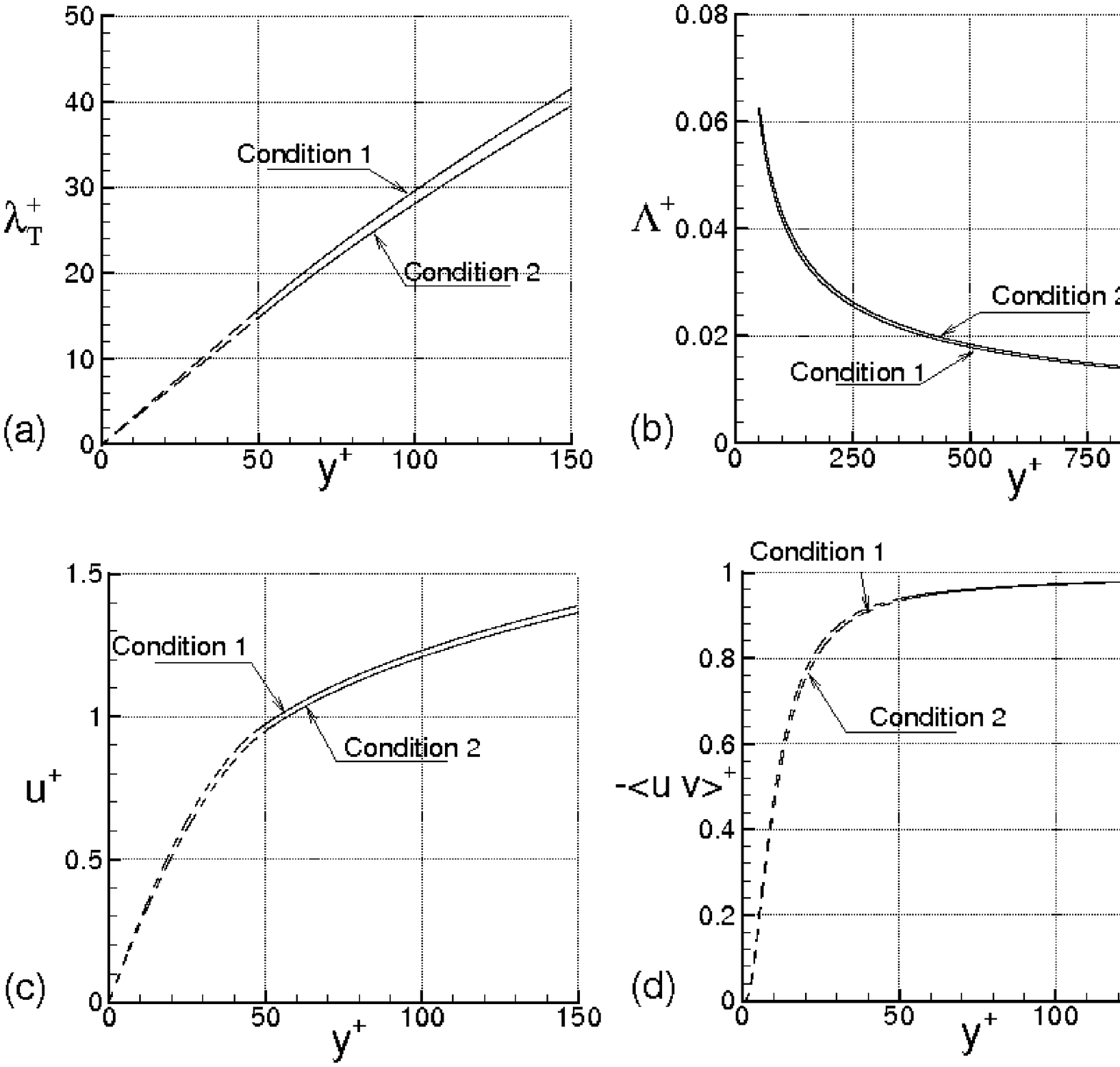}
\caption{Distribution of the dimensionless variables in the boundary layer: 
SL dashed lines,
TR continuous lines. 
(a) Taylor scale.
(b) Maximal Lyapunov exponent.
(c) r.m.s. of fluctuating velocity.
(d) Reynolds stress. }
\label{figura_6}
\end{figure}

The von K\'arm\'an constant and all the others parameters are shown in the table \ref{table1}.
 The table reports also the values of the free parameters $A$, $B$ and $C$ associated 
to the classical wall laws (\ref{wall law 00})-(\ref{wall law}), which are here calculated through the identification with  Eqs. (\ref{U turbo}) and (\ref{U ss}).
In particular, $C$ is calculated with Eq. (\ref{C}), whereas $A$, which pertains the velocity law in the buffer layer, is identified as the slope $d  U^+(y^+)/d \ln y^+$ of Eq. (\ref{U ss}) where $d^2 U^+(y^+)/{d \ln  y^+}^2 =0$, and $B$ is consequentely determined. 
The values of $k$ and $C$ are compared with the results given by different authors.

For both the conditions, $y^+_e \approx 50$ and $U^+_e \approx 15$ and this corresponds to a difference with respect to the classical data of \cite{Nikuradse} and \cite{Reichardt}, which is less than 2 $\%$.
As far as the Prandtl's length is concerned, according to Eq. (\ref{Prandtl}), it  varies quite similarly to $R_T$, and its derivative, calculated for $y^+ = y_e^+$, is slightly greater than $k$ 
(see Eq. (\ref{Prandtl_y}))
\bea 
\left( \frac{d l_p^+}{dy^+} \right)_e  =
1.34782... \left(  \frac{d \lambda_T^+}{dy^+} \right)_e  \simeq 1.00052 \ k
\eea
Its specific values, shown in the table, are in excellent agreement with the classical results. 
For what concerns $A$, $B$ and $C$, their values agree quite well the experiments, expecially for what concerns $C$ and $A$.

Once the free parameters are identified, $R_T$
is calculated in function of $y^+$ through  Eq. (\ref{R_T}) with
$O(y^+ -y^+_e)^2 \rightarrow 0$ and (\ref{ss}), for both the conditions (see Fig. \ref{figura_4}), and this corresponds to consider only the logarithmic profile of $U^+$.
Figure \ref{figura_5} shows $U^+(y^+)$ calculated with Eq. (\ref{U turbo}) and (\ref{U ss}). 
It is apparent that the results agree very well with the experimental data of different authors 
(\cite {Nikuradse}, \cite{Klebanoff_1}, \cite{Reichardt}) 
and with the formulas proposed by \cite{Spalding} and \cite{Musker}, with an error which does not never exceed 5 $\%$ in the interval of $y^+$ of the figure.
In particular, the condition 1 (Eq. (\ref{condizio 1})) provides a maximum difference
between present results and the data of Nikuradse and Reichardt, less than 3 $\%$ in
this interval.

The other variables are represented in Fig. \ref{figura_6}. 
The dimensionless Taylor scale (Fig. \ref{figura_6} a), linear in the laminar sub-layer and in the buffer region, remains a rising function of $y^+$ in the turbulent zone and exhibits, there, a slope which decreases slightly with $y^+$, whereas the Lyapunov exponent (Fig. \ref{figura_6} b), defined only in TR, is represented by a monotonically decreasing function of $y^+$.  This latter, being proportional to the square root of the turbulent dissipation rate ($\Lambda = u/\lambda_T$), agrees with the data of the different experiments (\cite{Fernholz}), at least where $U^+$ exhibits logarithmic profile.

\begin{figure}[t]
	\centering
 \hspace{-30.mm}        \includegraphics[width=0.60\textwidth]{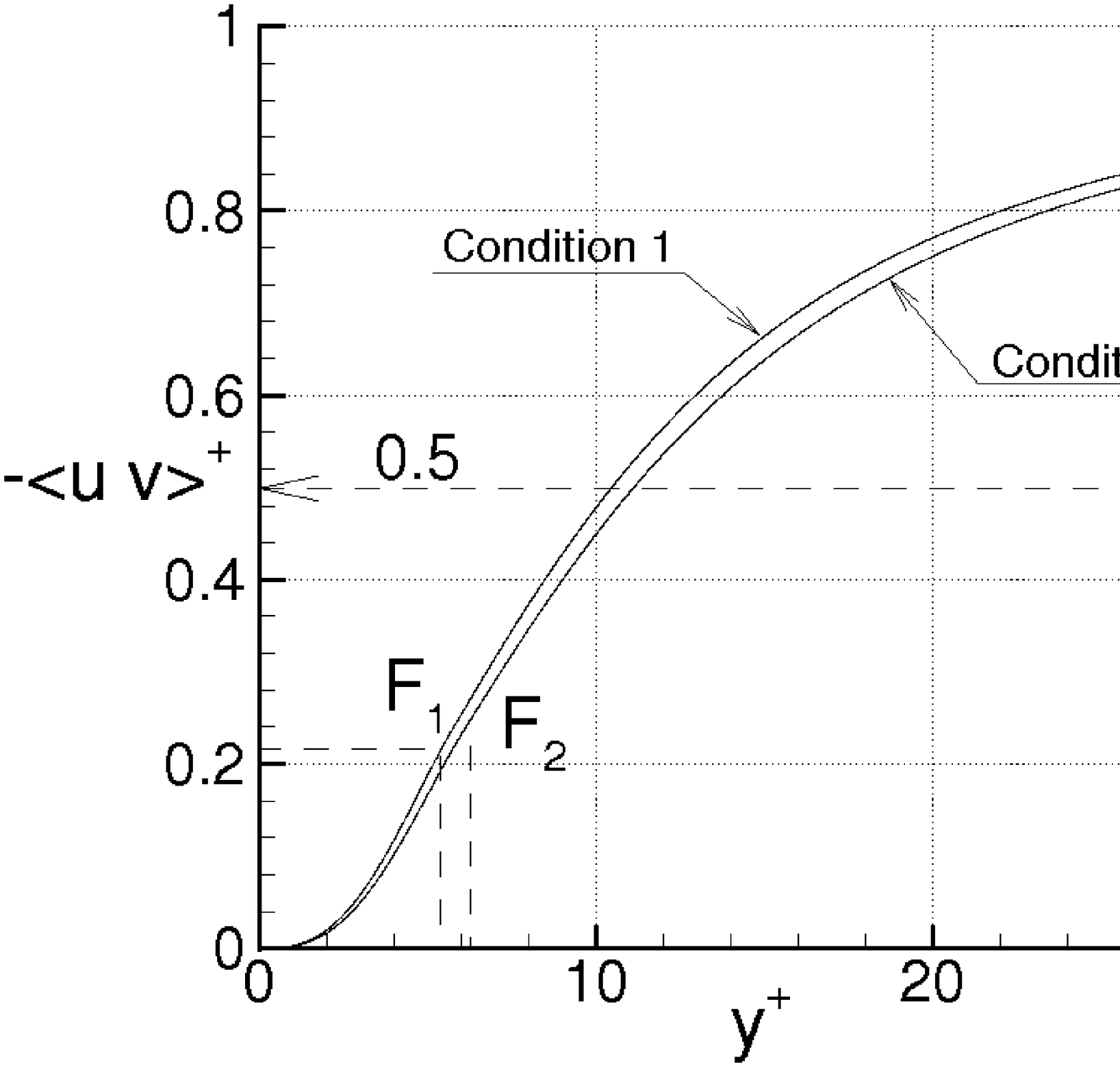}
\caption{Dimensionless Reynolds stress in the laminar sub-layer
and in the buffer region}
\label{figura_7}
\end{figure}
\begin{figure}[t]
\centering
 \hspace{-30.mm}        \includegraphics[width=0.60\textwidth]{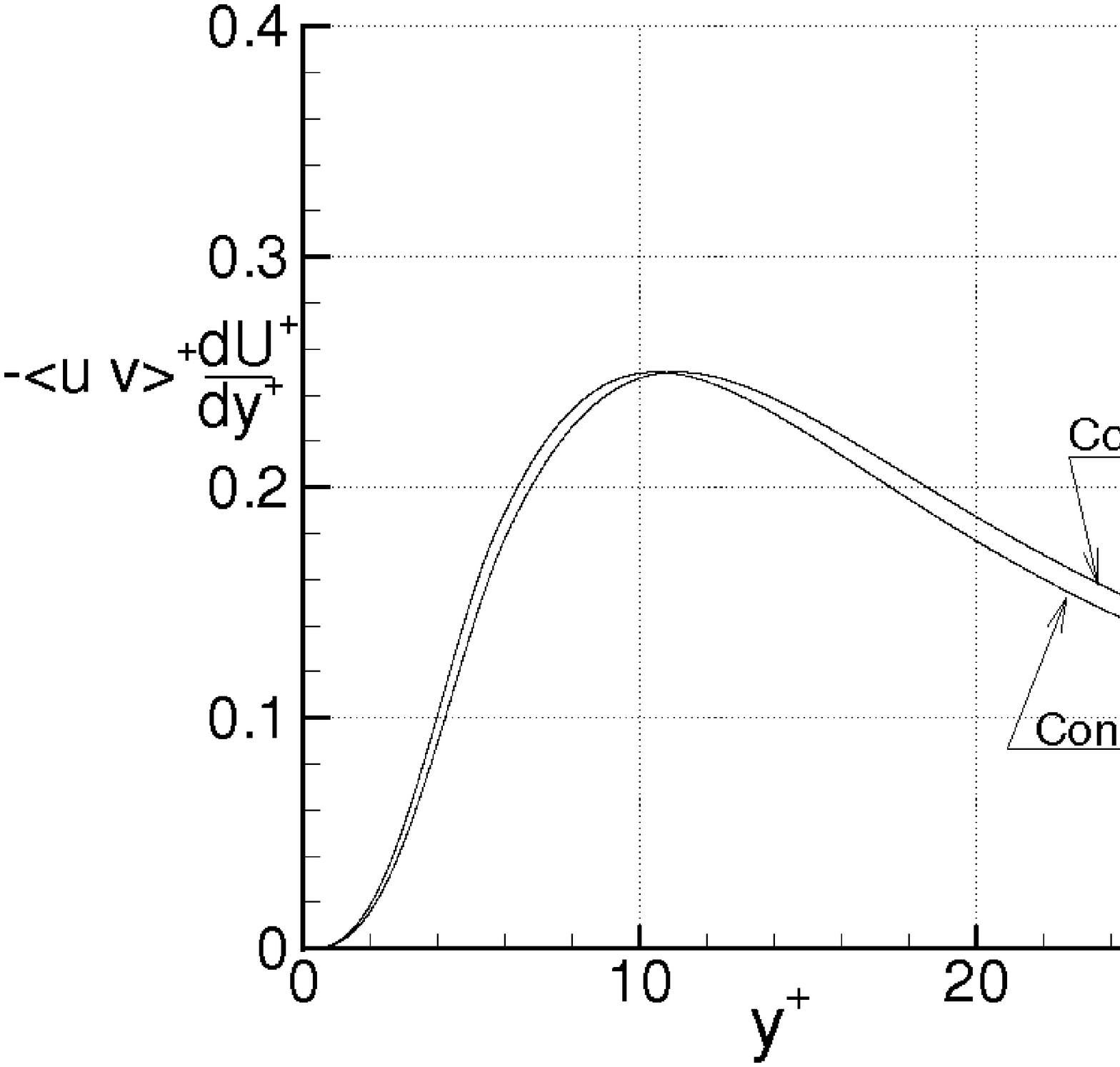}
\caption{Dimensionless rate of kinetic energy production in the laminar sub-layer
and in the buffer region}
\label{figura_8}
\end{figure}

The square root of the kinetic energy and the Reynolds stress are
shown in Fig. \ref{figura_6} (c) and (d).
These vary in TR following  
Eqs. (\ref{u})-(\ref{uv}) and are monotonically rising functions of $y^+$.
As seen, $u \approx y^+$ and $\langle u v\rangle^+ \approx {y^+}^3$ in the laminar sub-layer, 
whereas the Reynolds stress calculated with Eq. (\ref{uv}) in the buffer region, shows an 
inflection point $y_*^+$ which represents the separation element between LL and BL.
For $y^+> y_*^+$, $-\langle u v\rangle^+$ rises with $y^+$ until to reach the turbulent region where is about constant and equal to the unity.
More in detail, Fig. \ref{figura_7} shows $-\langle u v\rangle^+$ in an enlarged region which includes LL and part of BL. The distance $y_*^+$, represented by $F_1$ and $F_2$ is $y_*^+ =$ 5.734 and 6.097, in line with the order of magnitude of the laminar sub-layer tickness, and this is achieved at about $-\langle u v \rangle^+ =$ 0.24 in the two cases.
Next, the value $-\langle u v \rangle^+$ = 0.5 is obtained in the buffer layer for $y^+ \simeq$ 10.5 and 11.2, values in very good agreement with the classical results (\cite{Tennekes}, \cite{Hinze}).
This last condition corresponds also to the maximum of the kinetic energy 
production in the buffer layer, as shown in Fig. \ref{figura_8}. In this figure, 
the variations of the rate of turbulent kinetic energy production due 
to the mean flow are shown in LL and BL.
The shape of the diagrams follows the classical data (\cite{Tennekes}, \cite{Hinze} and references therein), and the values of $\langle u v \rangle^+$,  
$\langle u v \rangle^+ dU^+/dy^+$ and of the corresponding $y^+$, are in excellent agreement with the data of \cite{Tennekes} and \cite{Hinze}.

It is worth to remark that, although the monotonic trend of $u$ and $\langle u v\rangle^+$ can contrast some experiments which can give non-monotonic variations of these variables, the values of such quantities and the corresponding $y^+$, are comparable with those of the several experiments (\cite{Fernholz}). This discrepancy can be due the fact that, here, the effects of non-homogeneity of 
$\langle u_y ( p/\rho + u_j u_j/2 ) \rangle$ on the kinetic energy equation are neglected in TR, and that the average velocity is analyzed only in the logarithmic range.

\begin{table}[b]
\caption{Velocity law parameters and comparison of the results. P.R. as for "Present Result". }
\hspace{-20. mm}
\begin{tabular}{cccccccc} 
Parameter &  P. R.      &    P.   R.   & & & \\ 
            &      Condition 1    &    Condition 2       &    {Nikuradse}   &
   {Zagarola}      &   {Barenblatt}   &    Smith &  Fernholz \\
          &  Eq. (\ref{condizio 1}) &  Eq. (\ref{condizio 2}) &   &
  &  &  &   \\
\hline
\hline \\
$\ds k$                                 & 0.4233  & 0.3984  & 0.4  & 0.41 & 0.425 & 0.461 & 0.4 \\\\ 
$C$                                     & 6.2445  & 6.4822  & 5.5  & 5.2  & 6.79 & 7.13 & 5.1 \\\\
$A$                                     & 5.2384  & 5.5652   &      &      &      &      &     \\\\
$B$                                     & -4.1226 &  -4.7166 &      &      &      &      &     \\\\
\hline
\hline \\
$\ds \frac{r_{z}^*}{\lambda_{T e}}$           & 3.18233 & 3.38085 &      &      &    &  & \\\\
$\ds \left(\frac{d \lambda_T^+}{d y^+}\right)_e$ & 0.31423 & 0.29578     &      &    &  &  \\\\
$\ds \left(\frac{d l_p^+}{d y^+}      \right)_e$ & 0.4235 & 0.3986     &      &    &  &  \\\\
$\ds y^+_e$                                   & 49.3005 & 52.3760 &      &      &    &  &  \\\\
$\ds u^+_e$                                   & 0.96825 & 0.96825 &      &      &    &  &  \\\\
$-\langle u v \rangle_e^+$                    & 0.93750 & 0.93750 &      &      &    &  & \\\\
$U^+_e$                                       & 14.8250 & 15.7498 &      &      &    &  & \\\\
\hline
 \end{tabular}
\label{table1}
\end{table}

\bigskip

\section{\bf  Conclusions}
This work analyzes the turbulent wall laws through the Lyapunov theory of finite scale. The results, valid for fully developed flow with moderate pressure gradient, 
are subjected to the hypothesis that in the turbulent region, the energy
production due to the average flow is balanced only by the dissipation rate, 
whereas the non-homogeneity of $\langle u_y ( p/\rho + u_j u_j/2 ) \rangle$ 
is neglected.

The free parameters of the velocity law, here theoretically calculated through the
statistical properties of the velocity correlation functions, and the wall laws, 
are in very good agreement with the literature. 
In particular: 
\begin{itemize}
\item The von K\'arm\'an constant, theoretically identified as 
$k = (d R_T/d y^+)_e\approx 0.4 \div 0.42$, depends on the scale of the spanwise velocity correlation and does not requires the assumption of the logarithmic velocity profile.
\item The average velocity law and the distributions of the other dimensionless quantities such as kinetic energy and Reynolds stress in the boundary layer, agree -or at least are comparable- with experiments and direct simulations.
\end{itemize}

These results, which represent a further application of the analysis presented
in \cite{deDivitiis_3}, show that the finite scale Lyapunov analysis  
can be an adequate theory to explain the wall turbulence.

\bigskip

\end{document}